\documentclass[pr,nofootinbib,twocolumn,showpacs,aps]{revtex4-1}
\usepackage{bm}
\usepackage{graphicx}
\usepackage{amssymb}
\usepackage{amsmath}
\usepackage{eufrak}
\usepackage{color}
\usepackage[utf8]{inputenc}
\usepackage[unicode=true,colorlinks=true,urlcolor=blue,citecolor=blue]{hyperref}
\usepackage{pifont}
\usepackage{ulem}

\newcommand{\nix}[1]{}

\begin{document}

\title{Magneto-spatial dispersion of quantum wells}
\author{L.\,V.\, Kotova$^{1,2}$} 
\author{V. N. Kats$^1$} 
\author{A. V. Platonov$^1$} 
\author{V. P. Kochereshko$^1$} 
\author{R. Andr\'e$^3$} 
\author{L. E. Golub$^1$}
\affiliation{$^1$Ioffe Institute, 194021 St.~Petersburg, Russia}
\affiliation{$^2$ITMO University, 197101 St.~Petersburg, Russia}
\affiliation{$^3$Universit\'e Grenoble Alpes, CNRS, Institut NEEL, F-38000 Grenoble, France}
\begin{abstract}
Polarization conversion of light reflected from quantum wells governed by both magnetic field and light propagation direction is observed. We demonstrate that the polarization conversion is caused by the magneto-spatial dispersion in quantum wells which manifests itself in the reflection coefficient contribution bilinear in the in-plane components of the magnetic field and the light wavevector. The magneto-spatial dispersion is shown to arise due to structure inversion asymmetry of the quantum wells. The effect is resonantly enhanced in the vicinity of the heavy-hole exciton.  We show that microscopically the magneto-spatial dispersion is caused by the mixing of heavy- and light-hole states in the quantum well due to both orbital effect of the magnetic field and the in-plane hole motion. The degree of the structure inversion asymmetry is determined for GaAs/AlGaAs  and  CdTe quantum wells.
\end{abstract}

\maketitle{}

\section{Introduction}

Magneto-optical phenomena attract a great attention because they give an access to fundamental 
properties of studied media 
and due to perspectives of their use in quantum and optical information processing~\cite{Zvezdin_Kotov}. 
Magneto-optical properties of low-dimensional systems such as quantum wells (QWs) and other heterostructures are especially interesting due to their low symmetry allowing for transformation of light polarization in  transmission and reflection experiments.
All magneto-optical phenomena are useful to classify considering an expansion of the optical response functions over the magnetic field strength $\bm B$ and the light wavevector $\bm q$. 
For the susceptibility $\hat{\bm \chi}$ relating the dielectric polarization $\bm P$ and the electric field $\bm E$ as $\bm P = \hat{\bm \chi} \bm E$,
 we have to the lowest order in $\bm B$ and $\bm q$
\begin{equation}
\label{r}
\chi_{ij}(\bm B,\bm q) =  \chi_{ij}^0 + S_{ijk}B_k +  i\gamma_{ijk}q_k + C_{ijkl}B_kq_l.
\end{equation}
Here the first term describes 
birefringence, and the next contribution given by the tensor $\hat{\bm S}$ describes the well-known Faraday and magneto-optical Kerr effects. The term with the tensor $\hat{\bm \gamma}$ describes gyrotropic phenomena~\cite{Agranovich}. Nonzero components of $\hat{\bm \gamma}$ are allowed by symmetry only in systems lacking space inversion center. Moreover, 
some components of a vector and a pseudovector should belong to the same representation of the space symmetry group of the studied system. In QW structures, gyrotropy can be caused by both bulk and structure inversion asymmetry~\cite{LG_SG_review}. The effects caused by the linear in the wavevector terms in Eq.~\eqref{r} are special because they result in a difference of velocities of the circularly-polarized waves propagating in opposite directions. In particular, the tensor  $\hat{\bm \gamma}$ describes the optical activity of QWs demonstrated recently~\cite{Kotova}.

\begin{figure}[b]
\includegraphics[width=0.55\linewidth]{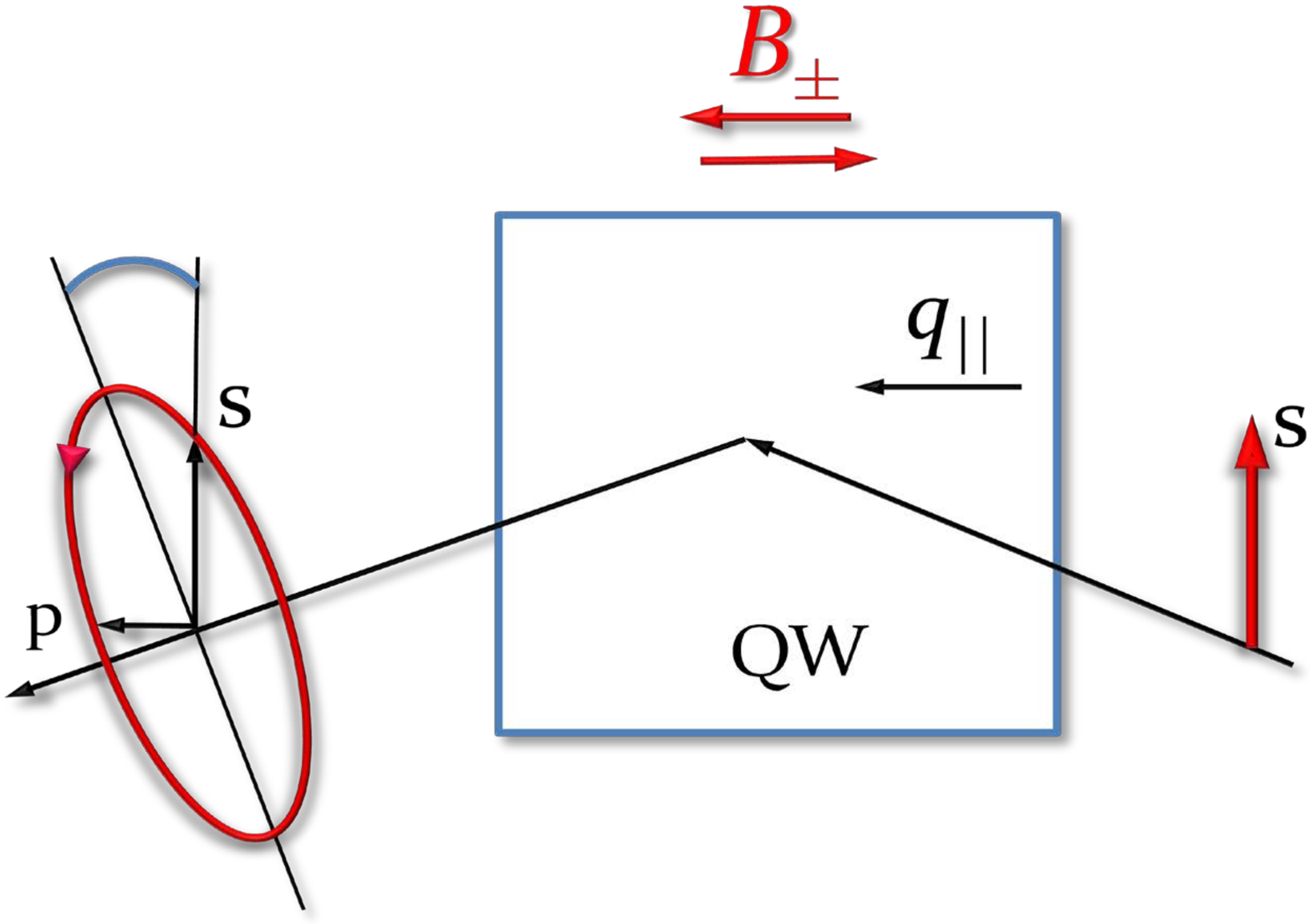}
\includegraphics[width=0.4\linewidth]{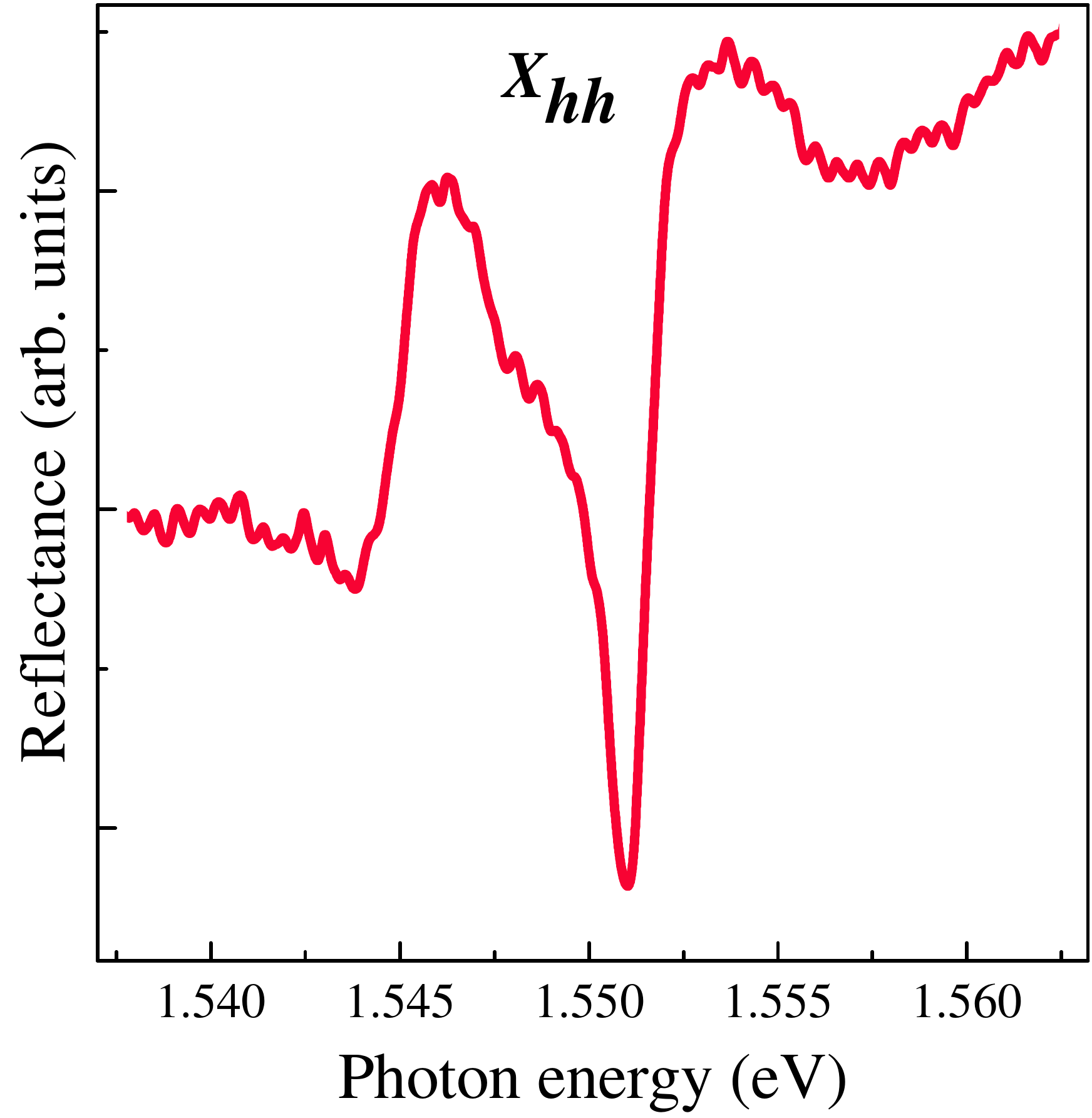}
\caption{(a): Experimental setup: $s$-polarized light at oblique incidence on a QW, magnetic field in the QW plane. The reflected light is elliptically polarized. (b) Reflectance spectrum measured from the GaAs asymmetric heterostructure in the vicinity of the $X_{hh}$ resonance at incidence angle $\theta_0=27^\circ$. }
\label{fig:Exp_scheme_Refl}
\end{figure}

The term with the tensor $\hat{\bm C}$ bilinear in both the light wavevector $\bm q$ and the magnetic field $\bm B$ describes the effect of magneto-spatial dispersion (MSD).
MSD is present in gyrotropic media only.
In contrast to the Faraday and magneto-optical Kerr effects, MSD is governed by the light propagation direction, and its symmetry is different from the pure  $\bm B$-linear part. For example, the terms linear in both $\bm q$ and $\bm B$ are invariant at time inversion operation while the $\bm B$-linear contributions change their sign. Therefore they contribute to different effects and can be distinguished by a proper choice of experimental geometry. 
MSD has been demonstrated in reflection and transmission experiments on various bulk semiconductors: GaAs~\cite{MSD_GaAs}, CdSe~\cite{MSD_CdSe}, ZnTe, CdTe~\cite{MSD_CdTe}, CdZnTe~\cite{MSD_CdZnTe}, thin films~\cite{MSD_thin films}, magnetic semiconductors~\cite{MSD_magn_semicond}, 
as well as in the second optical harmonics generation in ZnO~\cite{SHG_ZnO}.
Here we report on the observation of MSD in quantum well structures. 
It is well known that MSD effects are strongly enhanced in vicinity of exciton resonances~\cite{Agranovich,MSD_CdSe,MSD_CdZnTe,MSD_thin films}.
We  work in vicinity of the heavy-hole exciton resonance  in the geometry where the magnetic field $\bm B$ lies the plane of QW and the light incidence plane contains  $\bm B$, Fig.~\ref{fig:Exp_scheme_Refl}(a). It is well known that the Faraday and Kerr effects are absent at heavy-hole exciton for the in-plane magnetic field. Therefore observation of a $\bm B$-linear effect is a direct demonstration of  MSD.

MSD results in a polarization conversion of the reflected light. In particular, the $s$ or $p$ linearly polarized light is reflected elliptically polarized due to MSD, Fig.~\ref{fig:Exp_scheme_Refl}(a). The corresponding conversion coefficient linear in both $\bm B$ and $\bm q$ is observed for both III-V and II-VI asymmetrical quantum wells. 
We demonstrate that MSD exists due to structure inversion asymmetry (SIA) of the QWs under study.
The developed microscopic theory allowed us to explain quantitatively the amplitude of the MSD-induced reflection signal and to determine the SIA degree in the studied samples.

\section{Experiment}
\label{Exp}

For observation of the MSD effect, the sample should have a substantial degree of SIA. In QW structures, SIA can be introduced by two ways: i)~the heteropotential can have a triangular shape or 
ii)~left and right barriers of a rectangular QW have different heights.
In this work both possibilities are realized. A triangular GaAs/AlGaAs QW was grown by the molecular beam epitaxy (MBE) method on a semiinsulating substrate in the [001] direction. The structure contains a 200~nm wide Al$_{0.28}$Ga$_{0.72}$As barrier followed by the 8~nm wide QW. Then the other sloping barrier was grown 
with Al concentration smoothly increasing from 4~\% to 28~\% on a layer of width 27~nm. The structure design is identical to that of the sample used in Ref.~\cite{BIA_SIA_2006} where the electron spin-relaxation anisotropy was observed caused by a competition of the structure and bulk inversion asymmetries. It was demonstrated that such a structure design makes SIA strong enough for experimental observation of the Rashba spin-orbit splitting.
An example of a reflection spectrum measured on this sample in the vicinity of the heavy-hole exciton $X_{hh}$ is shown in Fig.~\ref{fig:Exp_scheme_Refl}(b).
The exciton resonance is clearly seen.  A rather high linewidth $\sim 5$~meV is caused by variations of the Al content during the QW growth process.

The rectangular 8~nm wide CdTe QW with different  barriers was grown on a Cd$_{0.96}$Zn$_{0.04}$Te substrate in the [001] direction. 
The left barrier is a 90~nm wide Cd$_{0.9}$Zn$_{0.1}$Te layer, 
and the right barrier is a 90~nm wide Cd$_{0.4}$Mg$_{0.6}$Te layer.

\begin{figure}[t]
\includegraphics[width=0.7\linewidth]{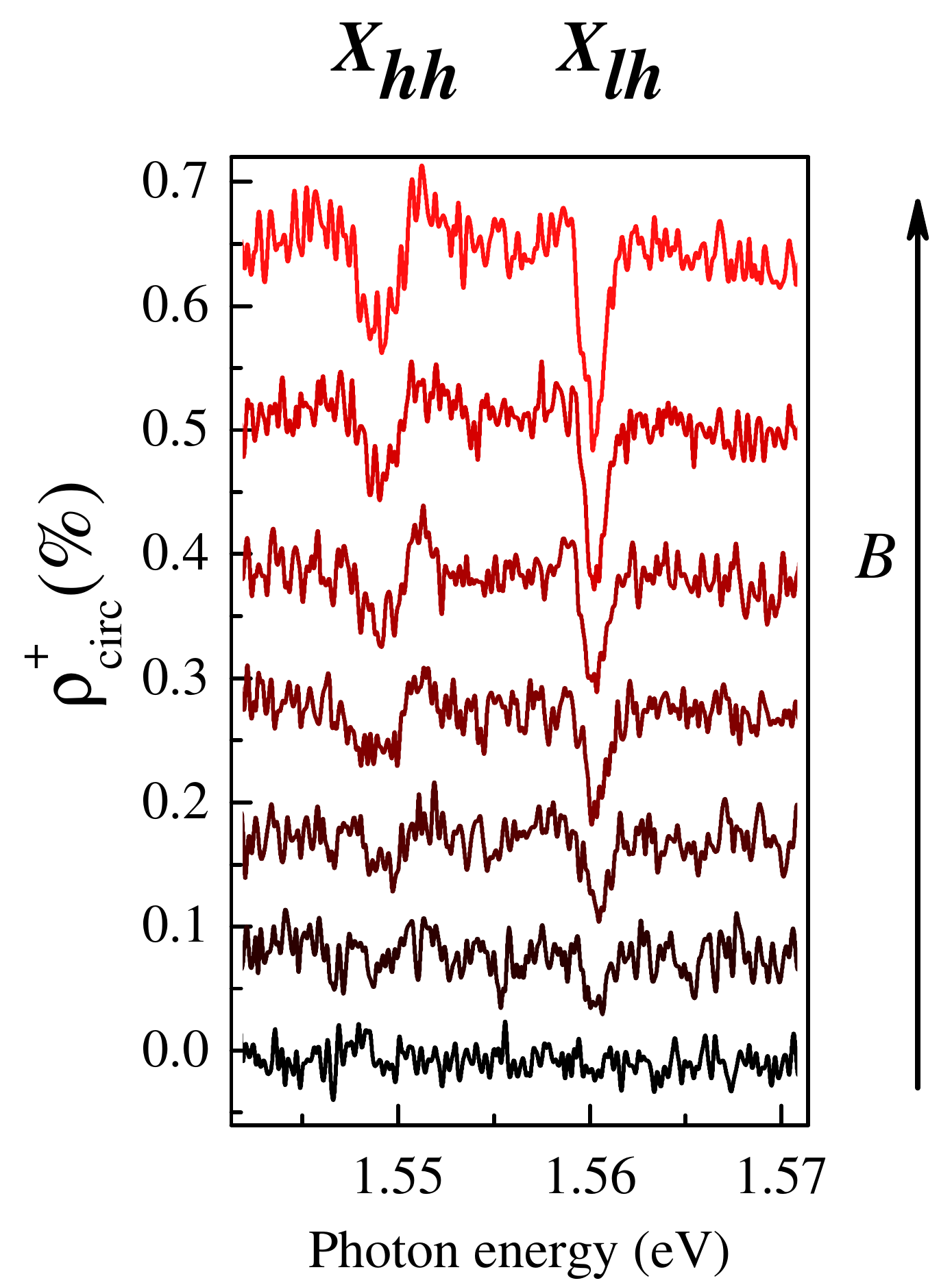}
\caption{Odd in $\bm B$ part of circular polarization of light reflected from the GaAs/AlGaAs asymmetric heterostructure.  The heavy-hole ($X_{hh}$) and light-hole ($X_{lh}$) exciton resonances are shown. Magnetic field $B=0.05, 0.2, 0.4, 0.5, 0.6, 0.8$ and 1~T (from bottom to top, the spectra are vertically shifted for clarity).  Temperature $T=3$~K.}
\label{fig:Waterfall_GaAs}
\end{figure}

We measured polarization of the light
reflected from both QW structures at oblique incidence. 
The incident light from a halogen lamp was linearly polarized in the plane of incidence ($p$ polarization) and perpendicular to it ($s$ polarization). 
This approach has been used 
recently to study the optical activity of QWs~\cite{Kotova}.
The experimental geometry is shown in Fig.~\ref{fig:Exp_scheme_Refl}(a). The magnetic field was produced by the electromagnet with a ferromagnetic core. This allowed us to have a magnetic field up to 1~T. 
The closed cycle helium cryostat was placed in the core gap. The measurements were performed at temperature $T=3$~K. The cryostat and electromagnet geometry limited the maximum incidence angle of light which in our case was $\theta_0=27^\circ$. All oblique incidence measurements were done at this angle. We have checked that the signal was absent at normal incidence. Spectral dependencies of the reflected light intensity $I(\omega)$ were registered by a CCD detector conjugated to a monochromator. 
Four polarization components of the reflected light ($I_{\sigma^\pm}$, $I_{\pm 45^\circ}$) were measured in magnetic fields from $-1$~T to $+1$~T. Here $I_{\sigma^\pm}$ are the intensities of the reflected light in right and left circular polarizations, and $I_{\pm 45^\circ}$ are the intensities of light linearly polarized at the angle $\pm 45^\circ$ to the  plane of incidence.

A direct analysis of the polarization spectra is difficult due to effects not related with MSD, e.g.  birefringence. Influence of these effects are stronger than the magnetic field induced polarization conversion. For an identification of MSD effect we used the fact that it is odd in $\bm B$ while the other contributions are either even or independent of the magnetic field. This allowed us to use for the analysis the differential signal measured for a fixed experimental setup and two opposite directions of the magnetic field. We analyzed the following values
\begin{equation}
\label{rho_circ_pm}
\rho_\text{circ}^\pm(\bm B) = {I_{\sigma^\pm} (\bm B)- I_{\sigma^\pm} (-\bm B) \over I_{\sigma^\pm} (\bm B)+ I_{\sigma^\pm} (-\bm B)},
\end{equation}
\begin{equation}
\label{rho_lin_pm}
\rho_\text{lin}^\pm(\bm B)= {I_{\pm45^\circ} (\bm B)- I_{\pm45^\circ} (-\bm B) \over I_{\pm45^\circ} (\bm B) + I_{\pm45^\circ} (-\bm B)}.
\end{equation}

\begin{figure}[t]
\includegraphics[width=0.75\linewidth]{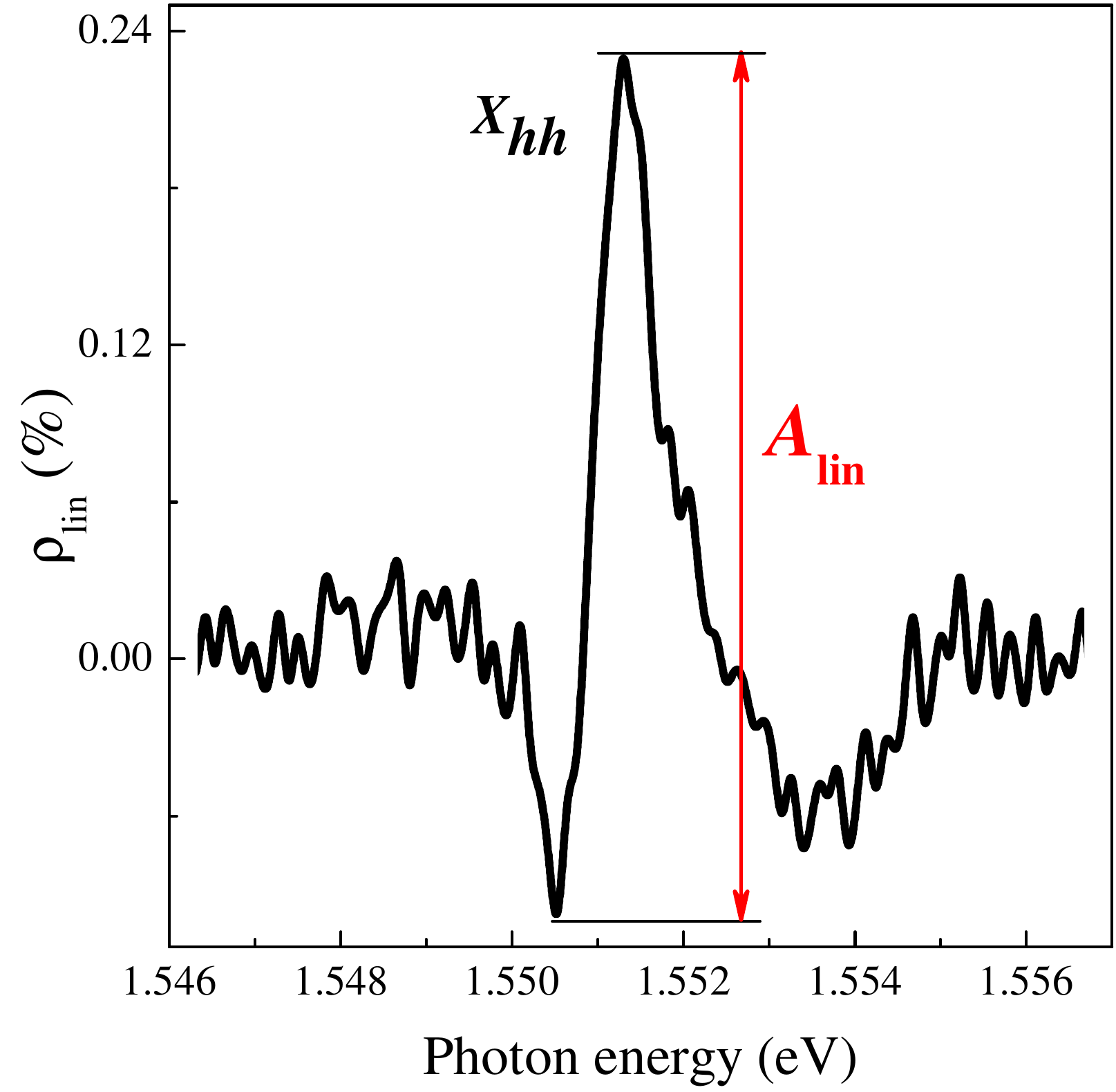}
\caption{Odd in $\bm B$ part of linear polarization of light reflected from GaAs/AlGaAs in the vicinity of the $X_{hh}$ resonance at magnetic field $B=1$~T. The arrow indicates the signal amplitude plotted in Fig.~\ref{fig:Flag_GaAs}.}
\label{fig:P_lin_GaAs}
\end{figure}

Figure~\ref{fig:Waterfall_GaAs} shows the spectral dependence of $\rho_\text{circ}^+(\bm B)$ for the GaAs triangular QW for magnetic fields from 0.05 to 1.0~T. The signal is absent in low fields. With increasing of the field, the signal arises for both $X_{hh}$ and $X_{lh}$ resonances being more pronounced for the latter. For $B=1$~T, $\rho_\text{circ}^+$ equals to 0.15~\% at the $X_{hh}$ line.

\begin{figure}[t]
\includegraphics[width=0.99\linewidth]{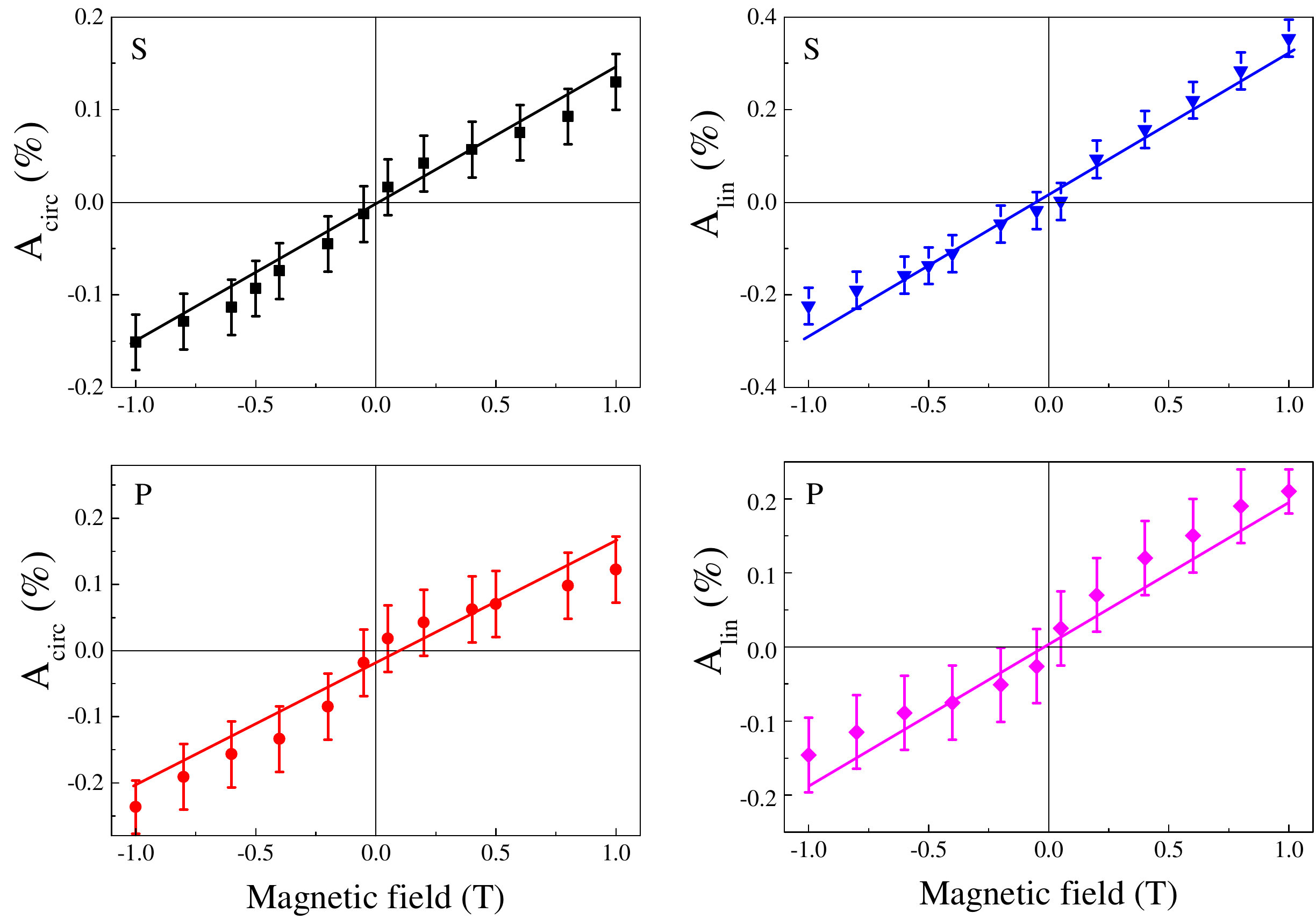}
\caption{Magnetic field dependencies of the signal amplitudes for GaAs/AlGaAs sample 
for circular and linear polarization at $s$ and $p$ polarizations of the incident light. }
\label{fig:Flag_GaAs}
\end{figure}

The four quantities given by Eqs.~\eqref{rho_circ_pm} and~\eqref{rho_lin_pm} are defined for the magnetic field of one sign. They can be combined to extend the domain of definition to the entire real axis, using the  relationship obvious for effects linear in the magnetic field
\begin{equation}
\label{plus_minus}
\rho_\text{circ, lin}^+(\bm B)=\rho_\text{circ, lin}^-(-\bm B).
\end{equation}
Experimentally Eq.~\eqref{plus_minus} means that by the simultaneous change of the field and polarization sign, the same odd in $\bm B$ value is measured. This allows us to introduce the value
\begin{equation}
\label{rho_circ}
\rho_\text{circ}(\bm B) = \left \{ 
\begin{array}{l}
\rho_\text{circ}^+(B), B>0,\\
\rho_\text{circ}^-(B), B<0,
\end{array}
\right.
\end{equation}
and similarly the value $\rho_\text{lin}(\bm B)$. The quantities $\rho_\text{circ}$ and $\rho_\text{lin}$ are experimentally obtained and analyzed below. 
At small values of the MSD signal, $\rho_\text{circ,lin}$ are obviously equal to the odd in $\bm B$ contributions to the Stokes parameters~\cite{Stokes_param}.

\begin{figure}[t]
\includegraphics[width=0.8\linewidth]{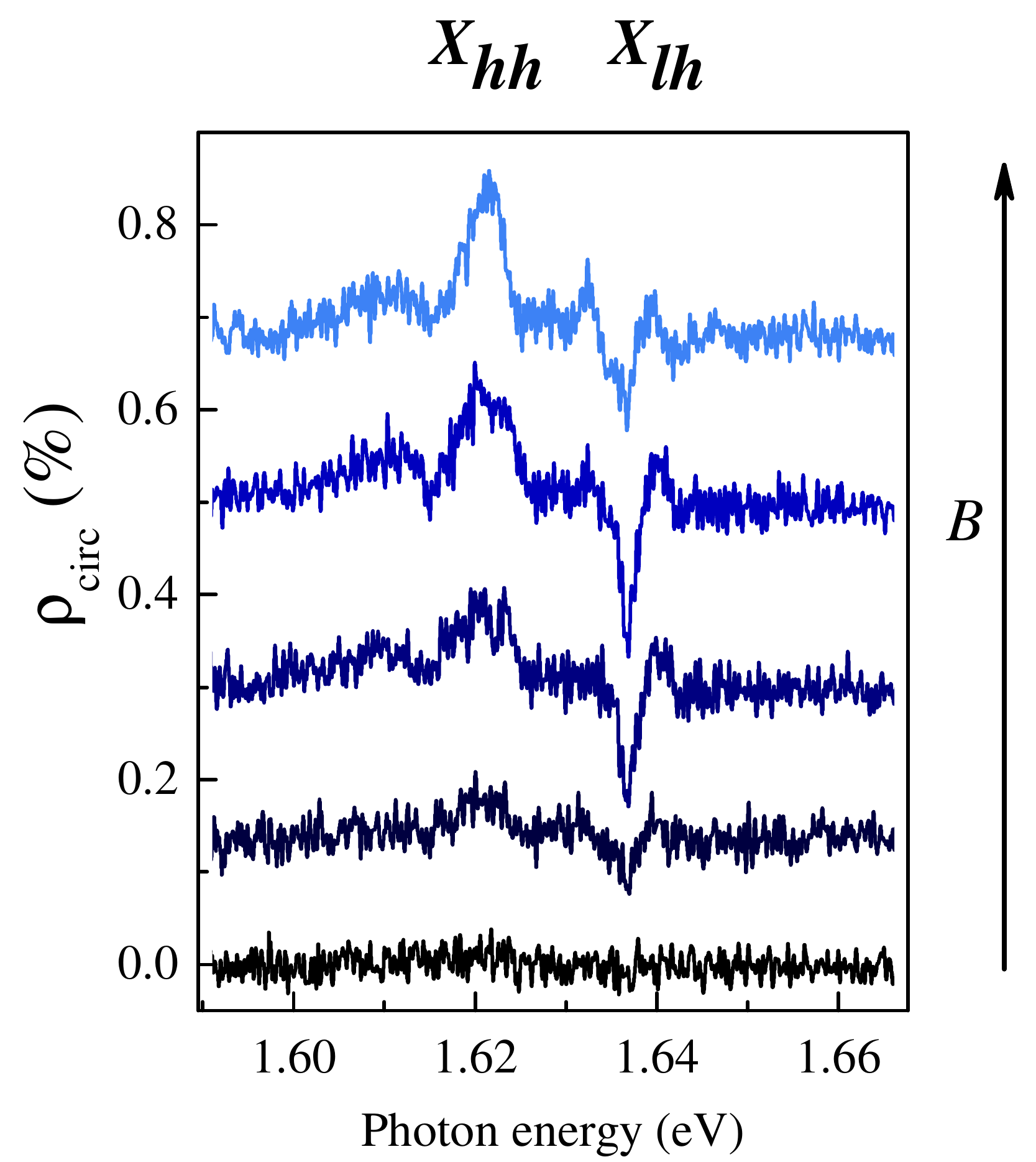}
\caption{Odd in $\bm B$ part of circular polarization of light reflected from CdTe sample. From bottom to top:  $B=0.05, 0.3, 0.5, 0.7$ and 0.95~T.}
\label{fig:Waterfall_CdTe}
\end{figure}

\begin{figure}[b]
\includegraphics[width=0.99\linewidth]{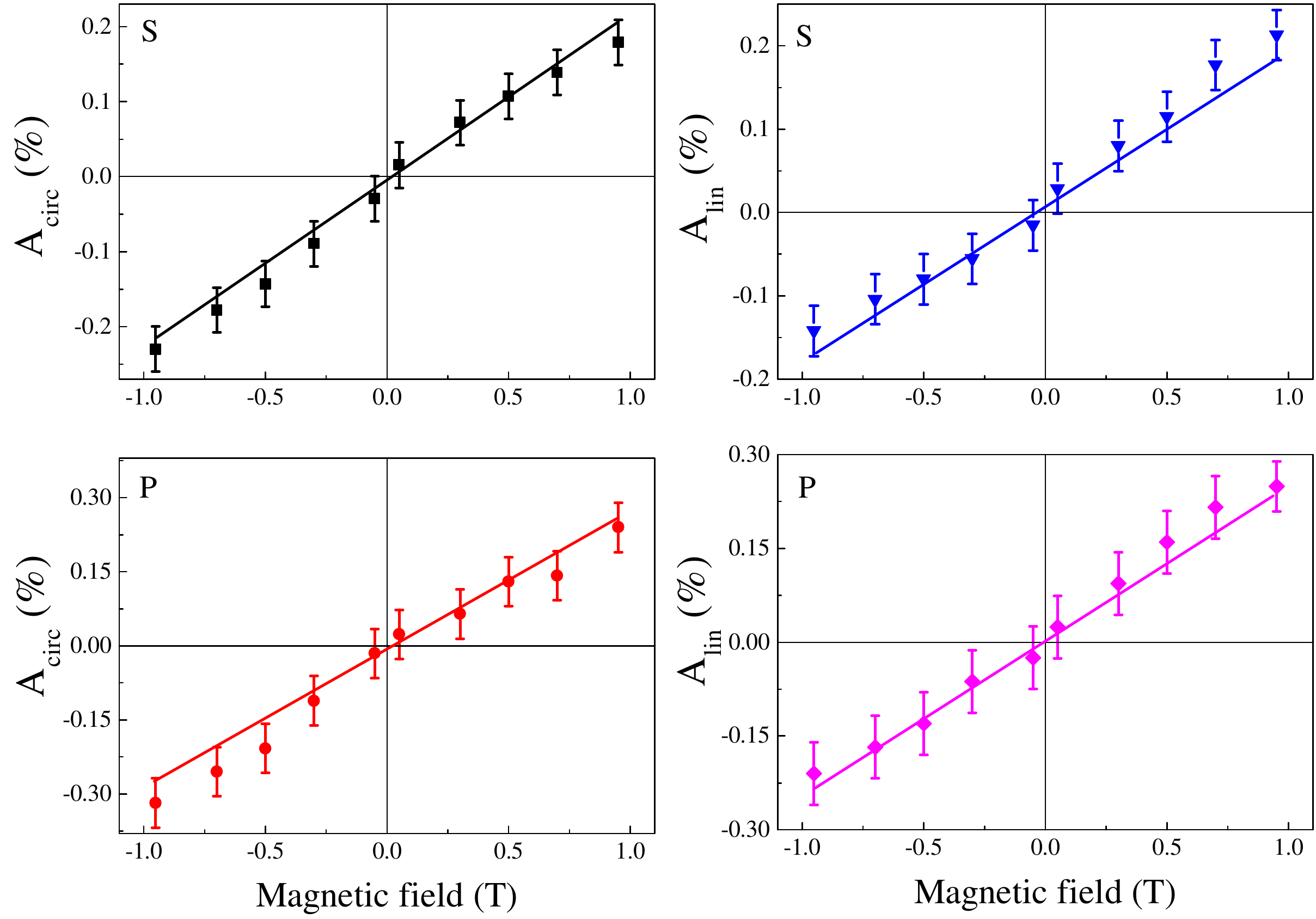}
\caption{Magnetic field dependencies of the amplitudes $A_\text{circ}$ and $A_\text{lin}$ in CdTe sample 
for circular and linear polarization at $s$ and $p$ polarizations of the incident light.}
\label{fig:Flag_CdTe}
\end{figure}

Figure~\ref{fig:P_lin_GaAs} shows $\rho_\text{lin}$ at $B=1$~T for the GaAs sample. The arrow indicates the signal amplitude $A_\text{lin}$ which is used for the analysis of magnetic-field dependencies. 
The value $A_\text{circ}$ is defined in a similar way for  $\rho_\text{circ}$~\cite{foot1}.
The amplitudes $A_\text{circ,lin}$ are shown in Fig.~\ref{fig:Flag_GaAs} for $s$ and $p$ polarization of the  incident light.  With a reasonable accuracy all four magnetic-field dependences are linear. 
One can see that the dependencies for $s$ and $p$ incident polarizations are almost identical.

Experimental results for the CdTe QW show a similar behavior. Figure~\ref{fig:Waterfall_CdTe} demonstrates $\rho_\text{circ}^+$ spectrum in the vicinity of $X_{hh}$ and $X_{lh}$ resonances. Similarly to the GaAs QW, the signal increases from zero to 0.2~\% for $X_{hh}$ with increasing of the magnetic field from 0 to 0.95~T.
Figure~\ref{fig:Flag_CdTe} summarizes the data for $A_\text{circ}$ and $A_\text{lin}$ for the CdTe QW. The linear field dependence is clearly seen for the whole magnetic field range.

\section{Theory}

In the present work, we study reflection from QWs in the vicinity of exciton resonances. 
Generally, it is described by a nonlocal integral relation between the dielectric polarization $\bm P(z)$ and the electric field $\bm E(z')$ where $z$ is the coordinate normal to the QW plane~\cite{EL_book}. Therefore the susceptibility $\hat{\bm \chi}(z,z')$ in Eq.~\eqref{r} depends on both coordinates but, due to homogeneity in the QW plane, the expansion over powers of $\bm q_\parallel$ is possible. 
%
According to the experimental geometry, Fig.~\ref{fig:Exp_scheme_Refl}(a), where the magnetic field lies in both the QW plane and the incidence plane, $\bm B \parallel \bm q_\parallel$, we deal with the MSD effects caused by SIA. Phenomenologically, it means that we can consider an idealized system with the  point  symmetry group $C_{\infty v}$. The symmetry consideration yields for this point group 
that
the following 
components of the susceptibility tensor Eq.~\eqref{r} and bilinear combinations of in-plane components of $\bm B$ and $\bm q$ 
are transformed according to the same representation 
$E_2$ 
\begin{align}
\label{rr}
\chi_{xy}+\chi_{yx} = C(q_xB_x-q_yB_y), \\ 
\chi_{xx}-\chi_{yy} = C(q_xB_y+q_yB_x). \nonumber
\end{align}
%
Here  $x,y$ are arbitrary axes in the QW plane, 
and $C$ is the single linearly independent component of the 4-rank tensor $\hat{\bm C}$ introduced in Eq.~\eqref{r}. 
Below we consider two microscopic mechanisms of MSD in structure-asymmetric QWs and calculate the value of the
polarization conversion coefficient.

\subsection{Orbital mechanism}

The heavy- and light-hole states in QWs are described in the model of the Luttinger Hamiltonian~\cite{EL_book}. Treating its off-diagonal terms as a perturbation, we obtain the wavefunction of the heavy hole in the ground size-quantized subband $hh1$ with the wavevector $\bm k$ in the QW plane in the following form
\begin{equation}
	\Psi_{hh1, \bm k}^{B=0} =  C_{hh1}(z) u_{3/2}
	+  {\gamma_2\hbar^2 k_+^2 \over m_0}  \sum_n {\phi_{lhn}(z) \over E_{hh1}-E_{lhn}}  u_{-1/2}.
\end{equation}
Here 
$k_+ = k_x + ik_y$, $\gamma_2$ is the Luttinger parameter, $C_{hh1}$ and $\phi_{lhn}$ are the functions of size quantization of the corresponding levels of the heavy and light holes (in the absence of SIA, the functions $C_{hh1}$ and $\phi_{lh,(2k+1)}$ are even, and $\phi_{lh,2k}$ are odd relative to the center of the QW), 
and $u_{\mu}$ are the Bloch amplitudes
\begin{equation}
	u_{3/2} = - {X+iY\over \sqrt{2}}\uparrow, \quad
	 u_{-1/2}=  {X-iY\over \sqrt{6}}\uparrow + \sqrt{2\over 3} Z\downarrow.
\end{equation}

In the magnetic field $\bm B$ lying in the QW plane the component $A_+(z)=-izB_+$ is nonzero.
Therefore making the Peierls substitution
\begin{equation}
	k_+^2 \to \left(k_+ -{e\over \hbar c}A_+ \right)^2 \approx k_+^2 - 2k_+ A_+(z),
\end{equation}
we obtain to the linear order in the field
\begin{align}
\Psi_{hh1, \bm k}^{\bm B} &=   C_{hh1}(z) u_{3/2}   \\
&- {2\gamma_2\hbar^2 \over m_0} k_+ \sum_n \phi_{lhn}(z) {  \left<lhn|A_+(z)|hh1\right>\over E_{hh1}-E_{lhn}}  u_{-1/2}. \nonumber
\end{align}
We choose the vector potential in the form $\bm A = (zB_y,-zB_x,0)$ where the point $z=0$ is taken in the center of the QW~\cite{foot2}. Then we obtain
\begin{equation}
\label{Psi_hh1_k}
	\Psi_{hh1, \bm k}^{\bm B} =  C_{hh1}(z) u_{3/2} + iB_+ k_+ F(z) u_{-1/2},
\end{equation}
where we omit the $\bm B$-independent quadratic in $k$ term,
\begin{equation}
	F(z) = {2e\gamma_2\hbar \over m_0c}  \sum_n \phi_{lhn}(z) {z_{lhn,hh1}\over E_{hh1}-
E_{lhn}},
\end{equation}
\begin{equation}
	z_{lhn,hh1} = \int\limits_{-\infty}^\infty dz \phi_{lhn}^*(z) \, z \, C_{hh1}(z).
\end{equation}
As a result, the state with the angular momentum 3/2 is active simultaneously in both $x$ and $y$ polarizations at $k \ne 0$:
\begin{align}
\Psi_{hh1, \bm k}^{\bm B} =  &	 - {1\over \sqrt{2}}\uparrow \left[ X C_{hh1}(z) 
		- {Y\over \sqrt{3}} B_+k_+ F(z) \right] \\
&	- {i\over \sqrt{2}}\uparrow \left[ Y C_{hh1}(z) 
		- {X\over \sqrt{3}}B_+k_+ F(z)\right] \nonumber \\
&		+ i \sqrt{2\over 3} B_+k_+ F(z) Z\downarrow. \nonumber
\end{align}
As a result, we obtain the terms linear in both $\bm B$  and $\bm k$ in the components of the dipole momentum density, $d_i = \left<e1\uparrow|\hat{d}_i|3/2\right>$
\begin{align}
\label{d}
	d_x  = - {ep_{cv}\over \sqrt{2} \omega_0 m_0} \left<e1|hh1\right> (1-i\xi B_+k_+), 
	\\
	d_y = - i {e p_{cv}\over \sqrt{2}\omega_0 m_0} \left<e1|hh1\right> (1+i\xi B_+k_+),
\end{align}
where $p_{cv}$ is the interband momentum matrix element.
The small real parameter $\xi$ which is nonzero due to SIA is given by 
\begin{equation}
\label{xi}
	\xi = 
	{2e\gamma_2\hbar \over \sqrt{3} m_0 c }  \sum_n { z_{lhn,hh1}\over E_{hh1}-E_{lhn}} 
 {\left<e1|lhn\right>\over \left<e1|hh1\right>}.
\end{equation}

The components of the exciton dielectric polarization satisfy the following equations~\cite{EL_book}
\begin{align}
&(\omega_0 - \omega-i\Gamma) P_x(z) \\
&= \Phi(z) \int dz' \Phi(z') \left[ |d_x|^2 E_x(z') + (d_xd_y^* + d_x^*d_y)E_y(z')\right], \nonumber\\
\label{P}
&(\omega_0 - \omega-i\Gamma) P_y(z) \\
&= \Phi(z) \int dz' \Phi(z') \left[ (d_xd_y^* + d_x^*d_y) E_x(z') + |d_y|^2 E_y(z')\right].\nonumber
\end{align}
Here the real function $\Phi(z)$ is the envelope function of the exciton size quantization at coinciding coordinates of electron and hole, $\omega_0$ and $\Gamma$ are the heavy-hole resonant frequency and linewidth, and we ignore a small difference of the resonant frequencies and linewidths of excitons with different polarizations. 
We obtain from Eqs.~\eqref{d}-\eqref{P} that
the MSD induced contribution to the susceptibility has the form 
of Eqs.~\eqref{rr},
with the nonlocal response function $C$ given by
\begin{equation}
C(z,z') =  {2\xi \nu\over \hbar} \left| {e p_{cv} \over \omega_0 m_0}\right|^2 \left| \left<e1|hh1\right>\right|^2  {\Phi(z)\Phi(z')\over \omega_0 - \omega-i\Gamma}.
\end{equation}
Here we take into account that the hole and the light wavevectors are related by
$\bm k=\nu \bm q_\parallel$,
where
\begin{equation}
\nu= {m_h \over m_e+m_h}
\end{equation}
with  $m_h$ and $m_e$  being the heavy-hole mass in the QW plane and the electron effective mass, respectively.

In the present work, we study light reflection from QWs. 
It is described by the reflection coefficient tensor $\hat{\bm r}$ relating the light fields of the incident ($\bm E^{0}$) and the reflected ($\bm E^{r}$) waves.
Solving the problem of light reflection~\cite{EL_book}, we obtain 
\begin{equation}
\left( 
\begin{array}{c}
E_s^r\\ E_p^r
\end{array}
\right)
= \left( 
\begin{array}{cc}
r_s + \Delta r_s & {\cal R}\\ {\cal R} &r_p + \Delta r_p
\end{array}
\right) \left( 
\begin{array}{c}
E_s^0\\ E_p^0
\end{array}
\right).
\end{equation}
Here $r_s$ and $r_p$
are given by 
the standard expressions for the reflection coefficients from the QW for $s$ and $p$ polarized light
\begin{equation}
\label{r_QW}
r_s^{QW}={i \Gamma^0_{s}\over \omega_0-\omega-i\Gamma}, \qquad
r_p^{QW}={i (\Gamma^0_{p}-\Gamma^0_\parallel)\over \omega_0-\omega-i\Gamma},
\end{equation}
where $\Gamma^0_{p,s}=\Gamma_0(\cos{\theta})^{\pm 1}$, $\Gamma^0_\parallel = 4\Gamma^0_p\tan^2{\theta}$ with $\Gamma_0$ being the exciton oscillator strength at normal incidence, and $\theta$ is the light propagation angle in the barrier material~\cite{EL_book}.
With account for SIA ($\xi \neq 0$), we get the MSD corrections to the reflection coefficients of $s$ and $p$ polarized light
\begin{equation}
\label{Delta_r}
	\Delta r_s = 
	-2 \xi \nu (q_xB_y+q_yB_x) r_s^{QW}, 
\quad	\Delta r_p = -\Delta r_s\cos^2{\theta},
\end{equation}
as well as the polarization conversion coefficient
\begin{equation}
\label{R}
	{\cal R} =  4\xi \nu(q_yB_y-q_xB_x)\cos{\theta} \: r_s^{QW}. 
\end{equation}

The polarization parameters of the reflected light observed in the experiment, Eq.~\eqref{rho_circ}, are given by~\cite{Kotova,Smirnov_Glazov}
\begin{equation}
\label{rho_via_R}
\rho_\text{circ} = 2\text{Im}\left({\cal R} / r_i \right), 
\quad
\rho_\text{lin} = 2\text{Re}\left({\cal R} / r_i \right),
\end{equation}
where $i=s,p$ is the incident light polarization, and $r_i$ is the reflection coefficient from the whole structure.

\subsection{Spin mechanism}

There is another microscopic mechanism resulting in the MSD effects in QWs. It is based on $\bm k$-linear mixing of the heavy- and light holes and  the Zeeman splitting of the light-hole states.
Let us take into account these two perturbations, ${\cal H}^{\bm k}$ and ${\cal H}^{\bm B}$.
The first mixes the states $3/2$ and $1/2$, and the second mixes the $1/2$ and $-1/2$ states:
\begin{equation}
	{\cal H}_{{1\over2},{3\over2}}^{\bm k} = {\gamma_3\sqrt{3}\hbar^2  \over m_0}k_zk_+, 
	\quad
	{\cal H}^{\bm B}_{-{1\over2},{1\over2}}= {e\hbar\over m_0 c} \varkappa B_+.
\end{equation}
Here $\gamma_3$  and $\varkappa$ are the Luttinger parameters~\cite{EL_book}. 
In the second order of perturbation theory, the wavefunction of the heavy-hole in the $hh1$ subband with the wavevector $\bm k$ has the form:
\begin{align}
&	\Psi_{hh1, \bm k}^{\bm B} =  C_{hh1}(z) u_{3/2} \\
&	+  {\gamma_3\sqrt{3} e \hbar^3 \over m_0^2c}   \varkappa k_+ B_+ \sum_n {\left<lhn |k_z|hh1\right> \over (E_{hh1}-E_{lhn})^2}  \phi_{lhn}(z) u_{-1/2}. \nonumber
\end{align}
We again obtained the hole wavefunction in the ground subband in the form Eq.~\eqref{Psi_hh1_k},
where for the spin mechanism
\begin{equation}
	F_s(z) = -{\gamma_3\sqrt{3} e \hbar^3 \over m_0^2c}  \varkappa \sum_n {\left<lhn |i k_z|hh1\right> \over (E_{hh1}-E_{lhn})^2}  \phi_{lhn}(z).
\end{equation}
Therefore the spin mechanism yields the above derived Eqs.~\eqref{Delta_r}-\eqref{rho_via_R} for the reflection coefficients where
\begin{equation}
\label{xi_s}
	\xi_s 
= 	-{e\gamma_3\hbar^3 \over m_0^2 c} \varkappa \sum_n {\left<lhn |i k_z|hh1\right> \over (E_{hh1}-E_{lhn})^2} 
 {\left<e1|lhn\right>\over \left<e1|hh1\right>}.
\end{equation}

\section{Discussion}

The MSD induced light polarization conversion is detected and analyzed in the vicinity of the heavy-hole exciton $X_{hh}$. It is crucial that the $X_{hh}$ state has zero dipole momentum along the growth direction $z$. Therefore an application of the in-plane magnetic field does not lead to the magneto-optical Kerr effect at $X_{hh}$. As a result, only account for a finite light wavevector allows explaining of the observed polarization conversion.

Theoretical consideration performed in the previous Section demonstrates that the polarization conversion spectrum $\cal R(\omega)$ has a resonance near the $X_{hh}$ exciton, Eqs.~\eqref{r_QW},~\eqref{R}. This resonant behavior is clearly seen in Figs.~\ref{fig:Waterfall_GaAs},~\ref{fig:P_lin_GaAs} and~\ref{fig:Waterfall_CdTe}.

The developed theory shows that the MSD contribution to the polarization conversion is controlled by SIA via the parameters $\xi$ and $\xi_s$, Eqs.~\eqref{xi},~\eqref{xi_s}. If the QW is structure-symmetric then for odd $n=1,3,5\ldots$ the coordinate and momentum matrix elements, $z_{lhn,hh1}$ and $\left<lhn |k_z|hh1\right>$, are zero while for even $n=2,4,\ldots$ the overlap of the electron and light-hole envelope functions $\left<e1|lhn\right>$ is zero. Therefore the parameters $\xi$ and $\xi_s$ are present in QWs with SIA only, and they change signs, for example, at a reversal of the normal electric field applied to a symmetric QW.

It follows from the developed microscopic theory that the orbital and spin  contributions to MSD are related by
\begin{equation}
	\left|{\xi_s\over\xi}\right| \sim \varkappa {m\over m_0},
\end{equation}
where $m$ has an order of the hole mass for the motion along the growth direction. Since for both GaAs and CdTe the product $\varkappa m/m_0 \approx 0.3$~\cite{Durnev_FTT}, the spin and orbital mechanisms give comparable contributions to MSD, but the orbital one is a little  stronger for both GaAs and CdTe QWs. 
We can estimate the SIA parameter as $\xi \sim (e/\hbar c)a^2 z_{lh1,hh1}$, where $a$ is the hole localization length in the QW structure, see Eq.~\eqref{xi}. For ${a=8}$~nm we get $\xi \sim (z_{lh1,hh1}/a)$~nmT$^{-1}$.
%
%
The MSD effect in reflection is the polarization conversion coefficient ${\cal R} \approx \xi q$. 
This yields for $q\sim 3\times 10^5$~cm$^{-1}$  an estimate ${\cal R}/B \sim 0.01(z_{lh1,hh1}/a)$T$^{-1}$.

Let us compare this theoretical estimate with experiment. Experimental results presented in Figs.~\ref{fig:Flag_GaAs} and~\ref{fig:Flag_CdTe} show the amplitude of MSD signal in both GaAs and CdTe-based QWs is $2\times 10^{-3}B$~T$^{-1}$. This corresponds to the 10~\% SIA degree in both QWs: $z_{lh1,hh1}/a \sim 0.1$. 
This is reasonable estimate for both structures. In the 8~nm wide CdTe QW under study, SIA is caused by the difference in the left and right barrier materials. This difference results in a shift of the hole wavefunctions from the point in the middle of the QW and, hence, to a nonzero matrix element $z_{lh1,hh1}$.
In the GaAs QW the right barrier height increases smoothly from zero to the left barrier height value at a width 27~nm, see Sec.~\ref{Exp}. As a result, both localization length $a$ and SIA degree are larger than in the CdTe QW. This explains the comparable polarization conversion coefficients in GaAs and CdTe QWs.

\section{Conclusion}

The magneto-spatial dispersion is demonstrated in both III-V and II-VI QWs. The MSD induced conversion of the light polarization state is observed in the magnetic fields $B\leq 1$~T.
The polarization conversion signal increases linearly with $B$ and reaches the values $2\times 10^{-3}$ at $B=1$~T in both studied samples. The developed theory based on the in-plane wavevector and magnetic-field induced mixing of the heavy- and light-hole states explains the measured signal assuming a reasonable 10~\% degree of SIA.
As an outlook, the geometry of in-plane magnetic field perpendicular to the incidence plane will be investigated. The corresponding studies allow to determine a degree of  the bulk inversion asymmetry which results in the magneto-spatial dispersion in this case.

\acknowledgments 
Optical measurements performed by  L.~V.~K. were supported by Russian Science Foundation (project 16-12-10503). 
L.~E.~G. thanks Russian Science Foundation (project 14-12-01067) for support of his theoretical work and ``BASIS'' foundation.


\begin{thebibliography}{99}
\bibitem{foot1} For the spectrum shown in Fig.~\ref{fig:P_lin_GaAs} we define $A_\text{lin}>0$. At an opposite phase of the spectrum $\rho_\text{lin,circ}(\omega)$ the amplitudes are defined with the negative sign.
\bibitem{foot2} Despite the wavefunction depends on the vector potential which can be shifted by an arbitrary constant $\bm A(z) \to \bm A(z)+ \bm A_0$, one can see that such a shift is equivalent to the substitution $\bm k \to \bm k - (e/\hbar c) \bm A_0$. This shift in the $\bm k$-space does not change measurable quantities.
\end{thebibliography}
\end{document}